\documentclass[12pt,english]{article}
\usepackage{babel}
\usepackage{amsmath}
\usepackage{amssymb}
\usepackage{natbib}

\def\pe#1 #2{\vec #1\cdot\vec #2}
\def\pes#1 #2 #3{\hbox{#1}\left(
\vec #2\right)\cdot\vec #3}

\def\eqref#1{(\ref{#1})}
\def\glosa#1{\marginpar{}}
\def\titulo#1{\vskip 0.3cm
\centerline{\sc #1}\vskip 0.3cm}

\def\sello{tracer mapping}
\def\RR{\mathbb{R}}
\def\g{\gamma}
\def\r{{\bf r}}
\def\rp{{\bf r}'}
\def\u{{\bf u}}
\def\up{{\bf u}'}
\def\R{{\bf R}}

\def\s{{\bf s}}
\def\v{{\bf v}}
\def\vp{{\bf v}'}
\def\V{{\bf V}}

\def\a{{\bf a}}
\def\ap{{\bf a}'}
\def\A{{\bf A}}

\def\comp{\circ}
\def\x{{\bf x}}
\def\y{{\bf y}}
\def\la{\lambda}
\def\nn{\nonumber}
\def\plano{T}
\def\recta{U}

\newtheorem{post}{Postulate}

\newenvironment{postulado}[1]{\begin{post}}{\end{post}}

\newtheorem{cor}{Corollary}

\newtheorem{defin}{Definition}

\newenvironment{definicion}[1]{\begin{defin}}{\end{defin}}

\newtheorem{theor}{Theorem}
\newenvironment{teorema}[1]{\begin{theor}}{\end{theor}}

\begin{document}

\title{Reference spaces in Special Relativity Theory: an intrinsic
         approach}

\author{Nilo C. Bobillo--Ares\thanks{ Departamento de Matem\'aticas,
Universidad de Oviedo, Facultad de Ciencias, C/ Calvo Sotelo s.  n.,
33007 Oviedo, SPAIN. E-mail: {\tt nilo@pinon.ccu.uniovi.es}}
      \and
Carlos Dehesa--Mart\'{\i}nez\thanks{ E.T.S.I. de Telecomunicaci\'on,
Universidad de
Valladolid, Campus ``Miguel Delibes", 47011 Valladolid, SPAIN. E-mail: {\tt
cardeh@gbien.tel.uva.es}}
}

\maketitle
\begin{abstract}
     Starting from a suggestion of Einstein on the construction of the
     concept of space, we elaborate an intrinsic method to obtain space
     and time transformations between two inertial spaces of reference,
     mathematically modeled as affine euclidean spaces.  The principal
     device introduced for relating the space readings in both
     spaces is the so-called {\it \sello}, which makes a snapshot
     of a space onto the other.  The general form of the space and time
     transformations is obtained
     as an affine--preserving mapping compatible with the principle of
     relativity, a cylindrical symmetry around the relative velocities
     between spaces and the group character of the transformations.
     After having obtained Galileo and Lorentz transformations, the same
     method has been applied to two classical problems: the Coriolis
     theorem of Newtonian Mechanics and the geometry of a rotating disk
     in Special Relativity.  Even in the case of Newtonian Mechanics,
     the possibility of distinguishing the spaces of reference is found
     useful.
\end{abstract}

\begin{itemize}

    \item[{\sc Key words}]: Concept of Space, Space and Time Transformations,
             Reference Frames.

\end{itemize}

\clearpage

\section{Introduction}
The nature of motion is an important issue since the first
stages in the science of mechanics.  As it happened with
many other topics, Galileo was
one of the first to deal with this problem in an
inquiring form. In his celebrated treatise {\it
Dialogue Concerning
the Two Chief World Systems Ptolemaic and Copernican},
\citet{amigo}, he faced the problem of the relativity of the
motion and argued that one
may refer a motion not only to the resting Earth but also to other
bodies moving with respect
to it.  Gradually, it was becoming clear that,
for studying a motion, use had to be made only of the space relative
to a body of reference,
making thus needless even to mention an absolute space. In spite
of this knowledge being so old, its operational meaning was
not fully understood until Einstein's findings on General
Relativity. In his book {\it The Meaning of Relativity}\/, \citet{meaning},
he set out clearly his ideas on the relativity of the space:
\begin{quotation}
``For the concept of space the following seems essential.
We can form new bodies by bringing bodies $B$, $C$, ... up to body
$A$; we say that  we {\it continue} body  $A$.  We can continue body
$A$ in such a way that it comes into contact with any  other body,
$X$.  The ensemble of all continuations of body $A$ we can designate
as the `space of the body $A$.' Then it is true that all bodies are in
the  `space of the (arbitrarily chosen) body $A$.' In this sense we
cannot speak of space in the abstract, but only of the `space
belonging to a body $A$.' The  earth's crust plays such a dominant
role in our daily life in judging the
relative positions of bodies that it has led to an abstract conception
of  space which certainly cannot be defended. In order to free
ourselves from this fatal error we shall speak only of `bodies of
reference,' or `space of reference.' "
\end{quotation}
According to Einstein, the space of a body is a physical concept and,
as such, operationally defined. He emphasized several features of it:
\begin{enumerate}
\item Any body may be used as a reference body. In special relativity
    we will limit ourselves to inertial bodies.
\item The points of a reference space have a permanent character to
the extent that they can be conceived as particles of some
continuation of the body. The points of a space of reference are
by definition at rest.
\item In a given space of reference, the record of the position of a
    moving particle is the point of that body which coincides with it at
    the considered time instant. There are no moving {\it points} but
    moving {\it particles} passing by close points in a space of
    reference, instead.
   \end{enumerate}

The fact that the space is relative in no way means that this concept
is irrelevant in physics. Energy or electric field
are also relative concepts but however we do not dismiss
them out as of lacking interest.
Even though this is clear enough, after the Minkowskian
geometric interpretation of spacetime the use of particular  bodies
of reference has been scorned as an old pre--relativistic prejudice
similar to absolute time or instantaneous action at a
distance. Nevertheless, to experimentally support a physical theory,
it has usually to recourse to data obtained in a certain laboratory
space. Therefore, it seems to be of the utmost interest to explicit
the usage rules for spaces of reference in the realm of special
relativity.

Besides the customary method based upon the use of
coordinates, the study of spacetime can be
undertaken in an axiomatic mathematically rigorous
way in which it is considered as an affine
pseudoeuclidean space. \citet{matolcsi} has
promoted such a procedure in order to build up an
intrinsic formalism upon which any physical theory can
be erected.
Our viewpoint is complementary: we are concerned
with the construction of spacetime from the  analysis
of the transformations between inertial reference
spaces.

Thus, the objectives this work is aimed to are:
\begin{enumerate}
\item To cast in mathematical form the idea of space of reference
    devised by Einstein.
\item Since the geometrical and physical relations on each body of
    reference do not depend on the chosen coordinate system, we will
    develop an intrinsic method, directly based on absolute objects of
    the Euclidean affine tridimensional space. The relativity of space
    demands to consider a
    different affine space associated to each body.
\item To present a theoretical device, that we call \sello, having a
    direct physical interpretation, suited to characterize the
    relationship between the physical quantities measured in different
    spaces, without coordinates. The concept of tracer is not new, but we
    think it has not received the attention it deserves. For instance,
    \citet{synge} introduced the term {\it snapshot}, taken from
    Milne's {\it world--map}, a term used in the same sense by \citet{rindler}.
\item To obtain {\it ab initio} intrinsic expressions of the Galileo
    and Lorentz transformations.
\end{enumerate}

The proposed method allows to wholly solve, or avoid, some little
problems appearing in
the usual derivations of space and time transformations, which are
really associated to a
particular choice of a coordinate system in each reference body. For
instance we could list: the problem of
parallelism between coordinate axes, addressed in \citet{frahm};
the reciprocity
relation for the relative motion of two inertial frames of reference,
analysed in detail in \citet{gorini}; and the search for Lorentz
transformations in cases with
arbitrarily oriented axes, carefully studied in  \citet{cushing}.

In order to state clearly the components used in the construction of
the space and time transformations, we have resorted to a style akin
to the ``more geometrico". However, this work is not addressed to a
search for a minimum set of postulates necessary for a logically closed
derivation [for that purpose see, for example,
\citet{matolcsi}, \citet{levy}, \citet{nishikawa},
\citet{schwartz} and References
cited therein]. Instead,
the intention of this work is of a methodological nature. We try to
clarify as far as possible the ingredients of physical character
which arise on occasion of the analysis of the problems concerning
the description of the measurements in space and time carried out
by several observers and the comparison between these measurements.

This paper is organized as follows.
In Section 2, using the \sello\ concept, the intrinsic method is
introduced. By adding further physical hypotheses, we obtain in
Sections 3 and 4 the transformations of Galileo and Lorentz. Next
Section is devoted to widen the scope of the intrinsic method by
giving an coordinate-free definition of angular velocity leading
to Coriolis theorem, within the limits of Newtonian mechanics, and a
characterization of the geometry on a rotating disk, in special
relativity.

\section{The intrinsic method}

\titulo{Spaces, times and events}

According to Einstein's construction, to every body $A$ there
corresponds a space of reference $K$. Besides, if the
body is an inertial one, its space is affine and euclidean. At every
point, the spaces are equipped with identical measuring rods, i. e.,
rods made by using the same instructions, which define the same length
unit in all of them.

From now on, the word ``space" as used in this paper will mean
the space of reference of a body. In that sense, a space has an
objective character and its points are permanent as they can be
considered as particles of a certain body.

Unless otherwise stated, only inertial bodies and spaces of reference
will be considered. For any of these spaces of reference, such as $K$,
being an affine
space, the set of its translations (pairs of points) forms a vector
space called $\cal V$. Following definitions and notation given in
  \citet{crampin}, Chap. 1,  we will represent a translation of
the point $Q$ by the vector ${\bf v}$ as a new point $P$ given by the
sum $Q+{\bf v}$.

On each space of reference identical clocks, which
define the same time measuring  unit in them, are distributed. If the
space is inertial, these clocks can be synchronized once and for all.

\begin{postulado}{Event manifold}
Each event $\cal P$ happens in the neighborhood of a single point $P$
of the space $K$. The clock at $P$ registers its date, $t$, a
single real number. The record $(t,P)$ fully characterizes the event.
\end{postulado}
In a similar way, the same event $\cal P$ is recorded in the space
$K'$  of other body as $(t', P')$. Two different records in  $K$
represent two different events that will produce also different
records in any other space. This circumstance confers the event
manifold an absolute character.

It is to be emphasized that $(t,P)$ is a very primitive way to
register an event. This record consists of a point and a real number
rather than four coordinates.

\titulo{Particles, trajectories and velocities}

\noindent A particle $M$ produces in the space $K$ the set of records
$(t,P_t)$,
$-\infty < t < +\infty$, called the trajectory of $M$, a curve in  $K$.

\begin{definicion}{Velocity}
The velocity $\bf v$ of $M$ with respect to $K$ at time $t$ is
\begin{equation}
{\bf v}(t)=\lim_{\Delta t\rightarrow 0}\frac{
\overrightarrow{P_tP_{t+\Delta t}}}{\Delta t}.
\end{equation}
\end{definicion}
Hence, ${\bf v}$ is a vector belonging to the space $\cal V$ associated
to $K$ and, thus, it is a relative quantity obtained from a series of
permanent records in $K$.

\titulo{Space and time transformations}

\noindent The study of the space and time transformations following
Einstein's guidelines can be undertaken by using the theoretical
devices introduced below.
\begin{definicion}{\sello\ }
The bijective mapping $\Lambda_t:K'\longrightarrow K$ maps each point
$P'\in K'$  onto the point $P\in K$ which mets $P'$ in the instant
$t$. We will refer $\Lambda_t$ to as {\em\sello}.
\end{definicion}
The suitability of this definition of $\Lambda_t$ as a bijective
mapping will be evident upon its explicit construction.
We call {\it trace} at $t$ of the point set
$\Omega'\subset K'$ on the space $K$ to the set
$\Omega=\Lambda_t(\Omega')$. The trace, being a set of points in $K$,
has a permanent character and it could be envisaged as a contact
photograph of the space $K'$. Also, note that, in order to get the
trace, no time consideration in $K'$ has to be made.
\begin{definicion}{Time at $K'$}
The event $\cal P$, characterized at $K$ by the pair $(t,P)$, happens
in $K'$ at the date $t'$. Therefore, we can introduce a function  $f'$
defined as
\begin{equation}\label{f_p}
t'=f'(t,P).
\end{equation}
The function $f'_P : \RR \to \RR $, defined by $f'_P(t)=f'(t,P)$ is
invertible.
\end{definicion}
The pair of relations
\begin{equation}\label{eq:directas}
\left\{
\begin{array}{rl}
P&=\Lambda_t(P'),\\
t'&=f'(t,P),
\end{array}
\right.
\end{equation}
will be called mixed space and time transformation formulas (note the
prime on those mappings with
image on $K'$). They
suffice for relating the records of any event  in the spaces $K$ and
$K'$. The mixed formulas are equivalent to the standard ones, in which
$(t',P')$ is obtained from  $(t,P)$. In fact, since $\Lambda_t$
has an inverse, call it $\left(\Lambda_t\right)^{-1}$, the first
of the Equations (\ref{eq:directas}) leads to
$P'=\left(\Lambda_t\right)^{-1}(P)$ which, together with
$t'=f'(t,P)$, constitutes the formulas for the standard
transformation. As we will see, the form (\ref{eq:directas})
will prove to be very suitable for our intrinsic derivation.
Moreover, it will be seen how certain symmetry considerations
are all we need to carry them to their final form.

From the invertibility of the mappings $\Lambda_t$ and
$f'_P$, stated in their respective definitions, it
follows that the mixed transformation formulas as a whole are
also invertible.
In fact, if $\left(\Lambda_t\right)^{-1}$ and
$\left(f'_P\right)^{-1}$ do exist, then
\begin{equation}\label{eq:inversomix}
\left\{
\begin{array}{rl}
P'&=\left(\Lambda_t\right)^{-1}(P),\\
t&=\left(f'_P\right)^{-1}(t'),
\end{array}
\right.
\end{equation}
allowing to write that
\begin{equation}
P'=\left(\Lambda_{\left(f'_P\right)^{-1}(t')}\right)^{-1}(P),
\end{equation}
which can be identified with the expression representing the trace
of $K$ on $K'$ at the time $t'$, $P'=\Lambda'_{t'}(P)$, i. e.,
with the first of the equations of the inverse mixed transformation.
Resorting in advance to the principle of relativity,
the spaces $K$ and $K'$
are on equal terms and, therefore, in the same way
as $\Lambda_{t}$, the \sello \ $\Lambda'_{t'}$ must
also be invertible, so that
\begin{equation}
P=\left(\Lambda'_{t'}\right)^{-1}(P').
\end{equation}
By substitution of this expression for $P$ in the second of the equations
(\ref{eq:inversomix}), we obtain
\begin{equation}
t=\left(f'_{\left(\Lambda'_{t'}\right)^{-1}(P')}\right)^{-1}(t'),
\end{equation}
which can be identified with the second of the equations for the
inverse mixed transformation, $t=f_{P'}(t')$.

Since the standard transformation formulas can be derived from the
mixed, the invertibility of the latter implies that of the former,
as it was to be expected.

\titulo{Spatial homogeneity of the transformation}

\noindent As previously stated, the inertial character of $K$ and $K'$
implies  that both spaces are homogeneous and isotropic. On the other
hand,  the transformation among spaces and times itself may not depend
on  the considered point. This property, which we will call spatial
homogeneity of
the transformation, is expressed as it follows:
\begin{postulado}{Homogeneity in space}\label{homo-espacio}
The mappings $\Lambda_t$ and $f'$ are homogeneous in their spaces,
i. e., they do not privilege any of their points.
\end{postulado}
Consider first the mapping  $\Lambda_t$. Since $K'$ is affine,
we have the obvious identity
\begin{equation}\label{hom-L}
\Lambda_t(Q'+\r')
=\Lambda_t(Q')+\overrightarrow{\Lambda_t(Q')\Lambda_t(Q'+\r')},
\end{equation}
where the vector
$\overrightarrow{\Lambda_t(Q')\Lambda_t(Q'+\r')}$, being a function of
$t$, $Q'$ and $\r'$, will be written as $\lambda_t(Q',\r')$. But
Postulate \ref{homo-espacio} means that a vector $\r' \in \cal V'$
placed at two different points of $K'$ must map into the same
vector in $\cal V$. Thus, $\lambda_t(Q',\r')=\lambda_t(R',\r')$, for
any pair of points $Q'$ and $R'$, and so the argument $Q'$ in
$\lambda_t$ can be suppressed. Instead of Eq. (\ref{hom-L}) we have,
then,
\begin{equation} \label{homoL}
\Lambda_t(Q'+\r') =\Lambda_t(Q') +\lambda_t(\r').
\end{equation}

Now, we can show that $\lambda_t$ is a linear mapping because of $K'$
being an affine space. In fact, consider two successive translations
in $K'$, $\r'$ and $\s'$, for which $(Q'+\r')+\s'=Q'+(\r'+\s')$, and
\begin{eqnarray*}
\Lambda_t\left[Q'+(\r'+\s')\right]&=&\Lambda_t(Q')
+\lambda_t(\r'+\s')\\
=\Lambda_t\left[(Q'+\r')+\s'\right]&=&\Lambda_t(Q'+\r') +\lambda_t(\s')\\
&=&\Lambda_t(Q') +\lambda_t(\r')+\lambda_t(\s').
\end{eqnarray*}
Thus, $\lambda_t(\r'+\s')=\lambda_t(\r')+\lambda_t(\s')$. Starting
with identities like
\begin{eqnarray*}
\lambda_t(2\r') = \lambda_t(\r'+\r')
=\lambda_t(\r')+\lambda_t(\r')=2\lambda_t(\r'),
\end{eqnarray*}
it is easily shown, for an $m$ being successively integer, rational
and real, that $\lambda_t(m\r')=m\lambda_t(\r')$, thus finishing the
proof.

With respect to $f'$, we can follow a similar line of reasoning. In
the obvious identity
\begin{eqnarray*}
f'(t,Q+\r)=f'(t,Q)+\left[f'(t,Q+\r)-f'(t,Q)\right]=f'(t,Q)+\theta'_t(Q,\r),
\end{eqnarray*}
the assumed spatial homogeneity in $f'$ implies that
$\theta'_t(Q,\r)$ must be independent on the particular point $Q$, then
\begin{equation} \label{dos}
f'(t,Q+\r)=f'(t,Q)+\theta'_t(\r).
\end{equation}
By combining two translations in $K$ we can show that $\theta'_t(\r)$
is linear. In fact,
\begin{eqnarray*}
f'\left[t,Q+(\r+\s)\right]=f'(t,Q)+\theta'_t(\r+\s)\\
f'\left[t,(Q+\r)+\s\right]=f'(t,Q+\r)+\theta'_t(\s)=
f'(t,Q)+\theta'_t(\r)+\theta'_t(\s),
\end{eqnarray*}
hence $\theta'_t(\r+\s)=\theta'_t(\r)+\theta'_t(\s)$.
And, finally, as in the case of $\lambda_t$,
$\theta'_t(m\r)=m\theta'_t(\r)$.

The same reasoning can now be used with the other argument in $f'$.
In the identity
\begin{eqnarray*}
f'(t+\tau,Q)=f'(t,Q)+\left[
f'(t+\tau,Q)-f'(t,Q)\right],
\end{eqnarray*}
the term between square brackets, in principle a function
of $t,\tau$ and $Q$, actually cannot depend on $Q$
according to the postulated homogeneity of $f'$,
therefore
\begin{eqnarray*}
f'(t+\tau,Q)=f'(t,Q)+g'(t,\tau),
\end{eqnarray*}
where $g'(t,\tau)$ depends on $t$ and linearly
on $\tau$, i. e., $g'(t,\tau)=\gamma'(t) \tau$. This last
statement can be shown by considering the
composition of two arbitrary translations in time. We get the relation
\begin{equation} \label{homof}
f'(t+\tau,Q+\r)=f'(t,Q) + \gamma'(t)\tau+\theta'_t(\r).
\end{equation}

\titulo{Time homogeneity of the transformation}
\noindent In a similar way as in the last Subsection, the
transformation  between spaces of reference may not depend on a
particular   chosen time. This property, that we will call time
homogeneity of  the transformation, is expressed as it follows:
\begin{postulado}{Homogeneidad temporal de las
transformaciones}\label{homo-tiempo}
The mappings $\Lambda_t$ and $f'$ are homogeneous in time,
i. e., they do not privilege any time instant.
\end{postulado}
An arbitrary point $P'$ of $K'$ describes in $K$ the
trajectory $\Lambda_t(P')$. The homogeneity in time of the mapping
$\Lambda_t$ implies that such a motion is uniform, i.e.,
\begin{equation}
\Lambda_t(P')=\Lambda_0(P')+\u(P')t.
\end{equation}
Here, $\u(P')$ is the velocity field linked to the particles
in $K'$, the {\it material velocity,} in the language of
the Kinematics of Continuous Media, see
   \citet{marsden}, p. 26. We may express this field also
in terms of the points of $K$
(usually called {\it spatial velocity}\/,
see \citet{marsden}, p. 27):
\begin{equation}
{\bf U}(t,P)=\u\left[\left(\Lambda_t\right)^{-1}(P)\right].
\end{equation}
Since time is homogeneous for the transformation,
$K$ must observe a steady situation in any point and, thus, the space
velocity field must be time independent:
\begin{equation}
\frac{\partial {\bf U}}{\partial t} =0.
\end{equation}
This condition can only be satisfied if $\u(P')$
is the same for every point $P'$.
It is worth noting that the relative motion between $K$ and $K'$
establishes a relationship between the homogeneities in time and
in space $K'$.

By differentiating (\ref{homoL}) with respect
to time we obtain the relation
\begin{equation}
\u(Q'+\r')=\u(Q')+\frac{d\lambda_t}{dt}(\r').
\end{equation}
If, as seen, every point of $K'$, in particular  $Q'$ and $Q'+\r'$,
have the same velocity, the following relation will hold
\begin{equation}
\frac{d\lambda_t}{dt}=0,
\end{equation}
that is, $\lambda_t$ must be time independent and, thus, it can be
represented simply as $\lambda$.

Two simultaneous events at $Q$ and $Q+\r$ happen in
$K'$ with a time separation that, according to Eq. (\ref{dos}), is
given by
\begin{equation}
f'(t,Q+\r)-f'(t,Q)=\theta'_t(\r).
\end{equation}
Time homogeneity in $f'$ requires that the above separation does not
depend on the chosen instant, $t$. Therefore, the index in
$\theta'_t$ is not necessary, so that
\begin{equation}
f'(t,Q+\r)=f'(t,Q)+\theta'(\r).
\end{equation}
On the other hand, for two events happening in the same point, $Q$,
with a time separation, $\tau$, we have, from Eq. (\ref{homof}),
\begin{equation}
f'(t+\tau,Q)-f'(t,Q)=\gamma'(t)\tau.
\end{equation}
In the same way, the condition for time homogeneity leads to $\gamma'$
to be independent of $t$.
The results concerning the last two Postulates can be summarized in the
following theorem:
\begin{teorema}{homogeneidad en el tiempo}
The homogeneities in space and time of the mappings
$\Lambda_t$ and  $f'$ imply
\begin{eqnarray}
\Lambda_t(Q'+\r') =\Lambda_0(Q') +\u t+\lambda(\r') \label{homoETL}\\
f'(t+\tau,Q+\r)=f'(t,Q)+\gamma'\tau+\theta'(\r)
\label{homoETf},
\end{eqnarray}
where $\u$ is the velocity of any particle in $K'$ as measured in $K$,
$\lambda$ is a linear mapping of  $\cal V'$ on $\cal V$,
$\gamma'$ is a number and $\theta'$ is a covector in $\cal V^*$.
\end{teorema}
In order to take advantage of the affine property of the spaces we
find useful to
use vectors to represent  points, which is carried out by choosing a
point and an instant as
an origin in each space and time. This process can be independently
made in each space but the
theory becomes simpler if one chooses a single, though arbitrary, event
$\cal Q$ as a reference or origin. This event registers at $K$ and
$K'$ as $(t_{\cal Q},Q)$ and $(t_{\cal Q}',Q')$,
respectively, and we have the relations
\begin{eqnarray}\label{referencia1}
\left\{
\begin{array}{rl}
Q&=\Lambda_{t_{\cal Q}}(Q'),\\
t_{\cal Q}'&=f'(t_{\cal Q},Q).
\end{array}
\right.
\end{eqnarray}
The choice of a reference event is the first step to coordinate the
space and time records in both spaces of reference.

An arbitrary event, $\cal P$, is registered as $(t_{\cal Q}+\tau,
Q+\r)$ and $(t'_{\cal Q} + \tau', Q'+\r')$ at $K$ and $K'$,
respectively. Hence, we can write
\begin{eqnarray}
\left\{
\begin{array}{rl}
Q+\r&=\Lambda_{t_{\cal Q}}(Q'+\r'),\\
t_{\cal Q}'+\tau'&=f'(t_{\cal Q}+\tau,Q+\r).
\end{array}
\right.
\end{eqnarray}
Then, by using Eqs. (\ref{homoETL}) and  (\ref{homoETf}),
the mixed transformation formulas  (\ref{eq:directas})
can be written in terms of vector quantities as
\begin{eqnarray}
{\bf r}={\bf u} \tau+\lambda({\bf r}')\label{erre}\\
\tau'=\gamma' \tau+\theta'({\bf r}).\label{taup}
\end{eqnarray}

\titulo{Some consequences of the principle of relativity}
\noindent The inverse transformation formulas,
\begin{equation} \label{eq:inversas}
\left\{
\begin{array}{rl}
P'&=\Lambda_{t'}'(P),\\
t&=f(t',P'),
\end{array}
\right.
\end{equation}
are merely those obtained by exchanging $K$ and $K'$. According to
the principle of relativity, both spaces of reference are physically
equivalent. Therefore we are allowed to repeat the previous analysis
but now from the point of view of $K'$.
Thus, we arrive to the vector relations, similar to (\ref{erre}) and
(\ref{taup}),
\begin{eqnarray}
{\bf r}'={\bf u}' \tau'+\lambda'({\bf r})\label{errep}\\
\tau=\gamma \tau'+\theta({\bf r}').\label{tau}
\end{eqnarray}
Here, $\up\in{\cal V'}$ represents the velocity of the particles of
$K$ as seen from $K'$.
\begin{teorema}{Relations of inversion}
The mappings and constants $\lambda$, $\lambda'$, $\theta$, $\theta'$,
$\gamma$  and  $\gamma'$ satisfy the identities
\begin{eqnarray}
{\bf u}+\gamma'\lambda({\bf u}')=0,&\label{eq:mapa1}\\
{\bf r}=\theta'({\bf r})\lambda({\bf u}')+\lambda\left[\lambda'({\bf
r})\right],&\quad\forall
{\bf r}\in {\cal V},\\
\gamma'\left[\gamma+\theta({\bf u}')\right]=1,&\\
\theta'({\bf r})+\gamma'\theta\left[\lambda'({\bf r})\right]=0,&\quad
\forall {\bf r}\in {\cal V},
\end{eqnarray}
and those resulting by interchanging the positions of primed and unprimed
symbols
\begin{eqnarray}
{\bf u}'+\gamma\lambda'({\bf u})=0,&\label{eq:mapa2}\\
{\bf r}'=\theta({\bf r}')\lambda'({\bf u})+\lambda'\left[\lambda({\bf
r}')\right],&\quad\forall
{\bf r}'\in {\cal V}',\\
\gamma\left[\gamma'+\theta'({\bf u})\right]=1,&\label{thp_u}\\
\theta({\bf r}')+
\gamma\theta'\left[\lambda({\bf
r}')\right]=0,&\quad \forall {\bf r}'\in {\cal V}'.
\end{eqnarray}
\end{teorema}
These results are obtained by imposing the condition that
(\ref{errep}) and (\ref{tau}) are the inverse ones of (\ref{erre}) and
(\ref{taup}), and then by substitution of the former into the later
ones.

\begin{postulado}{Measuring standards}
The following equalities hold
\begin{eqnarray}
\gamma'=\gamma, \label{eq:gamma}\\
u'=u,\label{modu}
\end{eqnarray}
where $u'=|{\bf u}'|$ and $u=|{\bf u}|$.
\end{postulado}
This statement presupposes that identical length and time measuring
standards have been adopted in both spaces, and it derives, then, from
the principle of relativity. We can make this connection plausible by
means of conceptual experiments like the following ones. In
relation to Eq. (\ref{eq:gamma}), consider the measurement of the
period of a standard clock, like a neutron lifetime, at rest in the
origin of $K$. Let $\Delta \tau_0$ and $\Delta \tau'$ be the values
obtained at $K$ and $K'$, respectively. Then, according to Eq.
(\ref{taup}), $\Delta \tau'=\gamma' \Delta \tau_0$. For the standard
clock now at rest in $K'$ one would obtain similarly $\Delta
\tau=\gamma \Delta \tau'_0$, where $\Delta \tau'_0$ and $\Delta \tau$
are the measured period in $K'$ and $K$, respectively. The equivalence
between  $K$ and $K'$, implicit in the principle of relativity,
implies (\ref{eq:gamma}); if not, a strange asymmetry between $K$ and
$K'$ would appear possibly  reminiscent of an ether at rest, and
both spaces could be distinguished from one another (see
  \citet{nishikawa} for a critical
analysis of these assumptions and their relationship to the principle
of relativity).
Besides, if the times in both spaces flow in the same direction, all
the time intervals considered so far are positive, and $\gamma>0$. A
similar argument involving the motion of the origins of both spaces
can be used to justify Eq. (\ref{modu}).

\titulo{Cylindrical symmetry of the transformation problem}
\begin{postulado}{Cylindrical symmetry}
In the transformation problem, at $K$, the direction $\recta$ of ${\bf
    u}$ is privileged.  All directions in the plane (subspace) $\plano$
    of vectors perpendicular to ${\bf u}$ are physically equivalent. In
    the same way, at $K'$, the direction $\recta'$ of ${\bf u}'$ is
    privileged and all directions in the plane $\plano'$ of vectors
    perpendicular to  ${\bf u}'$ are also physically equivalent.
\end{postulado}
\glosa{\begin{eqnarray*}
\lambda({\bf u}')=-\frac1{\gamma}{\bf u},\\
\lambda\left[\lambda'({\bf r})\right]={\bf r}
+ \frac1{\gamma}\theta'({\bf r}){\bf u},\\
\theta({\bf u}')=\frac{1-\gamma^2}{\gamma},\\
\theta\left[\lambda'({\bf r})\right]=-\frac{1}{\gamma}
\theta'({\bf r})
\end{eqnarray*}}
\glosa{\begin{eqnarray*}
\lambda'({\bf u})=-\frac1{\gamma}{\bf u}',\\
\lambda'\left[\lambda({\bf r}')\right]={\bf r}' +
\frac1{\gamma}\theta({\bf r}'){\bf u}',\\
\theta'({\bf u})=\frac{1-\gamma^2}{\gamma},\\
\theta'\left[\lambda({\bf r}')\right]=-\frac{1}
{\gamma}\theta({\bf r}').
\end{eqnarray*}}
This cylindrical symmetry implies that, if ${\bf r}_{\plano}\in T$,
then $\theta'({\bf r}_{\plano})$ must be independent on the vector
direction even though it will depend on its magnitude. To $\lambda$,
this symmetry demands that, if  ${\bf r}_{\plano}'\in T'$,
then $|\lambda({\bf r}_{\plano}')|^2$ is independent on the direction
of ${\bf r}_{\plano}'$. Stated otherwise, the image by
$\lambda$ of a circumference
of $T'$ has to be a circumference in $T$.
\begin{teorema}{ The covector $\theta'$}\label{teorematheta}
The action of the covector $\theta'$ on arbitrary vectors
is given by
\begin{equation}
\theta'({\bf r})=\frac{1-\gamma^2}{\gamma}\frac{{\bf r}\cdot{\bf
      u}}{u^2} \label{th_pr}.
\end{equation}
\end{teorema}
Proof to Equation (\ref{th_pr}): Given any vector ${\bf p}_{1}$ in the
plane $T$ we can add two more vectors in it, ${\bf p}_{2}$ and ${\bf
    p}_{3}$,  to form an equilateral triangle and such that
\begin{equation}
{\bf p}_{1}={\bf p}_{2}-{\bf p}_{3}.  \label{dif_p}
\end{equation}
Since the magnitude of the three vectors is the same, the action of
$\theta^{\prime}$ on any of them must be identical, i. e.,
\[
\theta ^{\prime }\left( {\bf p}_{1}\right) =\theta ^{\prime }\left(
{\bf p}_{2}\right) =\theta ^{\prime }\left( {\bf p}_{3}\right) =C.
\]
Applying $\theta ^{\prime }$ on both sides of Equation (\ref{dif_p})
and on account of its linearity, we conclude that $C=0$, that is, the
action of $\theta^{\prime }$ on any vector perpendicular to ${\bf u}$
is zero. This result allows us to fully characterize
$\theta^{\prime}$. In fact, let ${\bf r}_{\plano}$ and ${\bf
    r}_{\recta}$ be the
projections of a vector ${\bf r},$ ${\bf r}\in {\cal V}$, on the plane
$T $ and on the direction of ${\bf u}$, respectively.
As ${\bf r}_{\recta}=({\bf r}\cdot {\bf u}/u)({\bf u}/u)$
and $\theta ^{\prime }\left( {\bf r}_{\plano}\right)=0$,
then $\theta ^{\prime }\left( {\bf r}\right)=\theta ^{\prime}
\left( {\bf r}_{\recta}\right)$,
and by using Equations (\ref{thp_u}) and (\ref{eq:gamma}) we obtain
immediately Equation (\ref{th_pr}).
\begin{definicion}{Decomposition of $\lambda$}\label{deflambda}
We define the linear mappings
$\lambda_{\recta}:{\cal V}'\longrightarrow{\cal V}\,\,$ and
$\lambda_{\plano}:{\cal V}'\longrightarrow{\cal V}$ as
\begin{eqnarray}\nonumber
\left\{
\begin{array}{rl}
\lambda_{\recta}({\bf r}_{\plano}')&=0\\
\lambda_{\recta}({\bf r}_{\recta}')&=\lambda({\bf r}_{\recta}')
\end{array}
\right.;\hspace{0.1in}
\left\{
\begin{array}{rl}
\lambda_{\plano}({\bf r}_{\plano}')&=\lambda({\bf r}_{\plano}')\\
\lambda_{\plano}({\bf r}_{\recta}')&=0,
\end{array}
\right.
\end{eqnarray}
where ${\bf r}_{\plano}'\in T'$ and ${\bf r}_{\recta}'\in U'$.
\end{definicion}
These definitions are consistent since ${\cal V}'=U'\oplus T'$, and it is
straightforward to see that
$\lambda=\lambda_{\plano}+\lambda_{\recta}$. In the same way, by
introducing similar definitions for $\lambda'$, merely by exchanging
primed and unprimed letters, we write
$\lambda'=\lambda_{\plano}'+\lambda_{\recta}'$.
\begin{teorema}{The mapping $\lambda$}\label{thlambda}
The linear mapping $\lambda=\lambda_{\plano}+\lambda_{\recta}$ satisfies
the conditions
\begin{eqnarray}\label{lambdaU}
\lambda_U({\bf r}')&=&-\frac1{\gamma}\frac{{\bf r}'\cdot{\bf u}'}{u^2}{\bf
u}\\
{\bf r}_{\plano}'\in T'&\Rightarrow&\lambda({\bf r}_{\plano}')\in T\\
\lambda_T({\bf r}_{\plano1}')\cdot\lambda_T({\bf r}_{\plano2}')&=&
{\bf r}_{\plano1}'\cdot{\bf r}_{\plano2}';\qquad
{\bf r}_{\plano1}'{\mbox{, }}{\bf r}_{\plano2}'\in T'
\label{perpen}
\end{eqnarray}
\end{teorema}
Equation (\ref{lambdaU}) follows directly from Equation (\ref{eq:mapa1}) and
the result (\ref{eq:gamma}). The other ones  can be shown from the
cylindrical
symmetry of the transformation
and the principle of relativity, following a similar reasoning to the used
to prove Theorem \ref{teorematheta}.

Equation (\ref{perpen}) shows that $\lambda_T$ preserves the magnitude
of those vectors perpendicular to $\bf u'$ when transformed from $T'$
to $T$. As it is well known, a single parameter, usually an angle, is
enough to determine this transformation. The value of this angle is
so far undefined since we have not specified the orientation of  $K'$
with respect to $K$. Once the origins in spaces and times and the
directions of the velocities ${\bf u}$ and ${\bf u}'$ have been
established, the space $K'$ can be
rotated around ${\bf u}'$ without any change in the previous
results. From this consideration, it follows that no better
determination of  $\lambda$ can be obtained unless that missing piece
of information is added.

Finally, in order to fully characterize the mappings  $\theta'$ and
$\lambda_U$ we have to find out a relation between $\gamma$ and $\bf
u$.  To that aim, we will extend the method developed for
one-dimensional spaces in
\citet{levy} and accordingly we will assume that the space and time
transformations have a  group structure.  That group structure was
already explored in part when we assumed the existence of an inverse
transformation.

\titulo{\sc Closure of composition of transformations}
\noindent Let $K_1$, $K_2$ and $K_3$ be three inertial spaces of
reference.  Between each pair of them a space and time transformation
can be established:
$T^a$ between $K_1$ and $K_2$, $T^b$ between $K_2$ and $K_3$, and
$T^c$ between $K_3$ and $K_1$. We assume, as before, that a single
event defines origins in the three spaces and times. Each transformation
is characterized by a relative velocity
that can be measured in any of the two spaces related by the transformation.
Thus, $T^a$ is characterized by $\u^a_1$ and $\u^a_2$ in  $K_1$ and $K_2$,
respectively. Let $(\tau_1,\r_1)$ and $(\tau_2,\r_2)$
be records of the same event at $K_1$ and $K_2$, respectively. Our
previous analysis [refer to Eqs.  (\ref{erre}), (\ref{taup}),
(\ref{eq:gamma}),
(\ref{th_pr}) and (\ref{lambdaU}), and to Definition \ref{deflambda}]
allows us to write the equations of the mixed transformation
between those spaces as
\begin{equation}
          \r_1 = \la^a_{T1}(\r_2) + \u^a_1\tau_1 - \frac{1}{\g_a} \frac{\r_2
\cdot \u^a_2}{u^2_a}\u^a_1, \label{Ta1-1}
\end{equation}
\begin{equation}
\tau_2 = \g_a \tau_1 + \frac{1-\g^2_a}{\g_a}
\frac{\r_1 \cdot \u^a_1}{u^2_a}, \label{Ta1-2}
\end{equation}
where $u^2_a = \u^a_1 \cdot \u^a_1 = \u^a_2 \cdot \u^a_2$, $\g_a =
\g(u_a)$ and $\la^a_{T1}$ maps the perpendicular component (with
respect to
$\u^a_2$) of its $K_2$--argument onto a vector of $K_1$ perpendicular
to $\u^a_1$, preserving the length of that component. Equations
(\ref{Ta1-1}) and (\ref{Ta1-2}) must be considered on an
equal footing with the inverse equations obtained
interchanging indices 1 and 2:
\begin{equation}
          \r_2 = \la^a_{T2}(\r_1) + \u^a_2\tau_2 - \frac{1}{\g_a} \frac{\r_1
\cdot \u^a_1}{u^2_a}\u^a_2, \label{Ta2-1}
\end{equation}
\begin{equation}
\tau_1 = \g_a \tau_2 + \frac{1-\g^2_a}{\g_a}
\frac{\r_2 \cdot \u^a_2}{u^2_a}. \label{Ta2-2}
\end{equation}
The formulas for the other two transformations can be immediately deduced
from the preceding ones by changing on them the indices $a$, for the
transformation, and the related pair 1--2, for the linked spaces, to the
ones desired.

This set of equations will permit us to find an
universal form for the relation between the parameter $\g$ and the relative
velocity for any transformation as well as formulas relating the relative
velocities of the three transformations. For that purposes it is
easier to study the motion of the origin of each space from the other
two spaces together with the transformation linking the last ones.
According to the equations defining
the transformation $T^c$, the events experienced by the chosen
origin of $K_3$, $\r_3 = 0$, appear in $K_1$ as
\begin{equation}
\r_1 = \u^c_1 \tau_1, \label{r3-1a}
\end{equation}
\begin{equation}
\tau_1 = \g_c \tau_3. \label{r3-1b}
\end{equation}
In the same way, $\r_3= 0$ appears in $K_2$ as
\begin{equation}
\r_2 = \u^b_2 \tau_2, \label{r3-2a}
\end{equation}
\begin{equation}
\tau_2 = \g_b \tau_3,\label{r3-2b}
\end{equation}
obtained from the equations of the transformation $T^b$. The records
of such events registered in $K_1$ and $K_2$ are related by the
equations of the transformation $T^a$ given above. By substitution
of Eqs. (\ref{r3-1a})--(\ref{r3-2b}) into Eq.
(\ref{Ta1-1}) we get a relation among relative velocities:
\begin{equation}
\u^c_1 = \frac{\g_b}{\g_c} \la^a_{T1}(\u^b_2) +
\left[1-\frac{\g_b}{\g_a\g_c}\frac{\u^b_2 \cdot \u^a_2}{\u^2_a}\right]
\u^a_1. \label{u1c-u1a}
\end{equation}
Similarly, by eliminating times among Equations (\ref{Ta1-2}),
(\ref{r3-1b}) and (\ref{r3-2b}), the former becomes
\begin{equation}
\frac{\g_b}{\g_c} = \g_a + \frac{1-\g^2_a}{\g_a}\frac{\u^c_1 \cdot
\u^a_1}{\u^2_a}. \label{gb-gc}
\end{equation}

By studying in an identical way the motion of the origin of $K_2$ from
$K_1$ and $K_3$ and the equations of the transformation $T^c$ we arrive to
relations similar to the previous ones, namely
\begin{equation}
\u^a_1 = \frac{\g_b}{\g_a} \la^c_{T1}(\u^b_3) +
\left[1-\frac{\g_b}{\g_a\g_c}\frac{\u^b_3 \cdot \u^c_3}{\u^2_c}\right]
\u^c_1. \label{u1a-u1c}
\end{equation}
and
\begin{equation}
\frac{\g_b}{\g_a} = \g_c + \frac{1-\g^2_c}{\g_c}\frac{\u^c_1 \cdot
\u^a_1}{\u^2_c}. \label{gb-ga}
\end{equation}
Results similar to those given by Eqs. (\ref{u1c-u1a})--(\ref{gb-ga})
may be readily
obtained from the consideration of the motions
of the other possible pairs of origins.
Finally, by eliminating the product
$\u^c_1 \cdot\u^a_1$ between Eqs. (\ref{gb-gc}) and
(\ref{gb-ga}), and through analogous calculations
on the other mentioned results, we
arrive to the important identities
\begin{equation}
   \frac{1-\g^2_a}{\g^2_a} \frac{1}{u^2_a} =\frac{1-\g^2_b}
{\g^2_b} \frac{1}{u^2_b} =\frac{1-\g^2_c}{\g^2_c}
\frac{1}{u^2_c}. \label{id-ga-gc}
\end{equation}
Hence, the quantity $(1-\g^2)/(\g u)^2$ has a universal
value irrespective the
transformation it refers to. By denoting it
as $\alpha$, we get the advertised expression
for $\g(u)$:
\begin{equation}
    \label{eq:gam-u}
    \g=\frac{1}{\sqrt{1+\alpha u^2}}.
\end{equation}
The theory does not yield the actual value for $\alpha$
that will have to be obtained by
experimental means. The simpler choice is $\alpha=0$
leading to Galileo transformation,
briefly considered in Section \ref{sec:Galileo}. The choice $\alpha=-1/c^2$
corresponds to Lorentz transformation, studied in Section
\ref{sec:Lorentz}. The possibility of a positive value for
$\alpha$ can be discarded on causality arguments.
See \citet{levy} for more details.

\titulo{\sc Summary of space and time transformations}

\noindent For later use we collect here the results obtained in this
Section. The mixed transformation formulas between $K$ and $K'$,
transcribed  with an obvious
change in notation from  Eqs.  (\ref{Ta1-1}) and (\ref{Ta1-2}),  where
the  expression for $\g(u)$ given in Eq.
(\ref{eq:gam-u}) is used,  take the general form:
\begin{equation}
    \label{eq:mix-r}
    {\bf r'}=\lambda'_{\plano}({\bf r})+\left(\tau'-\frac1{\gamma}
\frac{{\bf r}\cdot{\bf u}}{u^2}\right){\bf u}',
\end{equation}
\begin{equation}
    \label{eq:mix-t}
    \tau=\gamma\tau'+\alpha \gamma {\bf r}'\cdot{\bf u}'.
\end{equation}
For the standard transformation, the corresponding formulas are
\begin{equation}
    \label{eq:est-r}
    {\bf r}'=\lambda'_{\plano}({\bf r})-\gamma
\frac{{\bf r}\cdot{\bf u}}{u^2}{\bf u}'+\gamma {\bf u}'\tau,
\end{equation}
\begin{equation}
    \label{eq:est-t}
    \tau'=\gamma\tau+\alpha \gamma {\bf r}\cdot{\bf u}.
\end{equation}

The important formula for velocity addition is easily obtained from
the (inverse of the) previous relations as:
\begin{equation}
    \label{eq:adi-vel}
    \v =\frac{\displaystyle\frac{1}{\gamma}
\lambda_{\plano}({\v}')+
\left(1-\frac{{\v}'\cdot{\u}'}{u^2}\right){\u}}
{1+\alpha \v'\cdot \u'},
\end{equation}
where ${\bf v}=d{\bf r}/d\tau$ and ${\bf v}'=d{\bf r}'/d\tau'$ are the
measured velocities of a particle in $K$ and $K'$, respectively.

These equations, together with the method used for obtaining them,
exhibit certain formal aspects which are worth to be emphasized.
\begin{enumerate}
\item Only intrinsic objects belonging to each space are used.
\item The mathematical formalism keeps each object separate in its own
    space. Thus, in Equation
(\ref{eq:adi-vel}) a vector of $K$ is written as the
sum of two vectors of the
    same space, and a numerical coefficient is obtained as the dot
    product of two vectors of the other space.
\end{enumerate}

\section{Galilean transformation. Absolute space}\label{sec:Galileo}

If one takes $\alpha=0$ then, irrespective of the value
of the relative velocity,
$\gamma=1$. Hence, from Eq. (\ref{eq:mix-t}), the
formula for time transformation is, simply,
\begin{equation}
\tau=\tau',\label{ttt}
\end{equation}
showing that the time elapsed between two events is he
same in any space of reference.  That is the old
concept of absolute time, implicit in Newton's Mechanics.
The transformation formula for position is, from Eq. (\ref{eq:mix-r}),
\begin{equation}
\label{galileo1}
    {\bf r'}=\lambda'_{\plano}({\bf r})-\left(
\frac{{\bf r}\cdot{\bf u}}{u^2}\right){\bf u}' +\u'\tau,
\end{equation}
which, together with Equation (\ref{ttt}), are the Galilean
transformation formulas between the spaces $K$
and $K'$. The classical law for velocity addition can now readily
obtained from Equation (\ref{eq:adi-vel}), with $\gamma=1$, as
\begin{equation}
\v = \lambda_{\plano}({\bf v}')+\left(
1-\frac{{\bf v}'\cdot{\bf u}'}{u^2}\right)\u.\label{com-v-gal}
\end{equation}

As it is easily shown, the linear mapping $\lambda$,
and not only $\lambda_T$, is an isometry, i. e.,
$\lambda({\bf r'})\cdot \lambda({\bf r'}) = {\bf r'}\cdot {\bf r'}$,
and thus defines a metric isomorphism between
the vectors in $\cal V'$ and $\cal V$. Therefore, a
metric isomorphism between $K'$ and $K$ can be
established which makes possible to
identify both spaces at any time according to the  rules:
\begin{enumerate}
\item Place $Q'$ on $\Lambda_t(Q')=Q+(t-t_0){\bf u}$;
\item Place ${\bf u}'$ on $\lambda({\bf u}')=-{\bf u}$; and
\item Place a vector ${\bf p}'\in \plano'$ on $\lambda({\bf p}')$.
\end{enumerate}
In this way, any point of $K'$ is identified with its trace in $K$.
The idea of an absolute space, as defined in \citet{desloge}, comes
from the possibility of
this identification. We should point out that this
identification is possible at any speed, even at
relative rest. This is the reason why figures showing two frames of
reference in relative motion seem so obvious: in
spite of the inherent limitations of a still picture forcing us to draw not
only $K$ but also $K^{\prime }$ at rest, such identification is always
allowed. On the other hand, since  this
identification is not possible in the spacetime of Minkowski when
$K^{\prime }$ is moving, such simple figures cannot properly be drawn,
situation that
we discuss in the following Section.

We may fuse $K$ and $K'$ into one space by setting
$\rp\equiv\lambda(\rp)$ and $-{\bf
u}\equiv\lambda({\bf u}')$ and obtain the well known transformation formula
\begin{eqnarray}
{\bf r}=\tau{\bf u}+{\bf r}'.\label{clasica}
\end{eqnarray}
This classical result entails a trap very difficult to escape
from. The Galilean transformation formula (\ref{clasica}),
an innocent vector
addition, seems to be a direct consequence of the affine character of the
(absolute) space. When it was shown that this was not an accurate physical
law it resulted rather natural to doubt the affine property of the
space by introducing an spatial anisotropy along the motion of the
body of reference. That idea was formulated as the FitzGerald
contraction hypothesis. If, on the contrary, we keep the separation
between spaces and write the Galilean transformation as
Equation (\ref{galileo1}),
consequence of an absolute time, the jump to special relativity is
easier since there is more room to look for a solution
elsewhere. Einstein solved the difficulty by questioning the absolute
character of time in his celebrated 1905 paper.

\section{Lorentz transformation}\label{sec:Lorentz}

By choosing a suitable negative value for $\alpha$ one arrives to
Lorentz transformation. Experimentally, one finds that, actually,
$\alpha=-1/c^2$, where $c$ is the speed of light in a
vacuum. Accordingly, the actual dependence of $\g$ on $u$ is
\begin{equation}
\g=\frac{\textstyle 1}{\textstyle \sqrt{1-\frac{\textstyle u^2}{\textstyle
c^2}}}. \label{VALORGAMMA}
\end{equation}
The formulas for the mixed Lorentz transformations are
Eqs.  (\ref{eq:mix-r}) and
(\ref{eq:mix-t}) with this value for $\g$, i. e.,
\begin{eqnarray}
\r&=&\la_{\plano}(\rp)- \left(
\frac{\textstyle 1}{\textstyle\g}\frac{\textstyle
\rp\cdot\up}{u^2}\right)\u +\u\tau, \label{LORMEZ1}\\
\tau'&=&\g \left(\tau-\frac{\textstyle
\r\cdot\u}{c^2}\right). \label{LORMEZ2}
\end{eqnarray}
The inverse transformation formulas are readily obtained by exchanging
primed and unprimed symbols in the later expressions.
From these equations we get the standard Lorentz transformation
\begin{eqnarray}
\r&=&\la_{\plano}(\rp)+ \g\left(\tau'-\frac{\textstyle
\rp\cdot\up}{u^2}\right)\u, \label{LORENTZ1}\\
\tau&=&\g \left(\tau'-\frac{\textstyle
\rp\cdot\up}{c^2}\right). \label{LORENTZ2}
\end{eqnarray}
Now, from Equation (\ref{LORMEZ1}), it can be seen that
\begin{equation}
\la(\rp)\cdot\la(\rp)=\rp\cdot\rp-\frac{\textstyle
(\rp\cdot\up)^2}{\textstyle c^2}, \label{NOORT}
\end{equation}
showing that $\la$ is not an isometric mapping. This fact has the
important consequence for the spaces $K$ and $K'$ that they cannot be
identified to each other. Therefore, figures mixing two spaces, even if
one-dimensional, are impossible, and only figures in spacetime can be
drawn if one insists in showing both spaces on them.

\section{Extending the method to two classical problems}
The intrinsic method based on the concept of  \sello\ may be used
in more general situations. As examples of that we have chosen two
classical problems, the analysis of which is,
in our opinion, more  transparent when
the proposed method is used.

\subsection{Intrinsic definition of angular velocity}

In Newtonian Mechanics, the spaces of two inertial bodies of reference,
$K$  and $K'$, may be metrically identified since the mapping $\lambda$
preserves the magnitude of vectors. This is true independently of the
relative velocity between the bodies. This fact makes possible to study
the transformations between two spaces of reference having {\it arbitrary
motions}\/. As before, $\Lambda_t:K'\rightarrow K$ is the \sello\ of $K'$ on
$K$.
\paragraph{Postulate} {\it In Newtonian mechanics, the \sello\
$\Lambda_t$ between two spaces $K$ and $K'$, having arbitrary relative
motion,
is affine and isometric.}

\vspace{0.1in}
\noindent We take as origins the arbitrary points $Q$ of $K$ and $Q'$ of
$K'$. As before, we write
the condition for the mapping $\Lambda_t$ to be affine, at any time, as
\begin{equation}
\Lambda_t(P')=\Lambda_t\left(Q'+\overrightarrow{Q'P'}\right)=
\Lambda_t(Q')+\lambda_t\left(\overrightarrow{Q'P'}\right), \label{afin}
\end{equation}
but now we cannot take $\lambda_t$ as time independent since
different points of $K'$  have different velocities.
Equation (\ref{afin}) in vector form is
\begin{equation}
\r=\R+\lambda_t\left(\rp\right)\label{directa}
\end{equation}
where
\begin{equation}
\r(t)=\overrightarrow{Q\Lambda_t(P')},\,
\R(t)=\overrightarrow{Q\Lambda_t(Q')} \mbox{ and }\,
\rp(t)=\overrightarrow{Q'P'}.
\end{equation}
By differentiating  (\ref{directa}) with respect to time, now
absolute,  we obtain a relation between velocities,
\begin{equation}
\v=\V+\dot \lambda_t(\rp)+\lambda_t(\vp(t)),\label{sum_v}
\end{equation}
where $\v(t)=\dot\r(t)$, $\V(t)=\dot\R(t)$ and $\vp(t)=\dot\rp(t)$.
From now on, the $t$ index in $\lambda$ will be suppressed for a
simpler notation.

Let us introduce the operator $\Omega : \cal V \rightarrow \cal V$
defined as
\begin{equation}
\Omega = \dot \lambda\comp\lambda^{-1}.
\end{equation}
By direct differentiation of the identities
$\lambda\comp\lambda^{-1}=\mathrm{id}$ and
$\lambda^{-1}(\x)\cdot\lambda^{-1}(\y)=\x\cdot\y$, we obtain the
relation:
$$
    \Omega(\x)\cdot\y=-\x\cdot\Omega(\y),
$$
showing that $\Omega$ is an antisymmetric operator.

In terms of this operator, Equation (\ref{sum_v}) can be
rewritten as
\begin{equation}
\v=\V+\Omega\left[\lambda(\rp)\right]+\lambda(\vp)\label{veloc}.
\end{equation}
Let $\boldsymbol{\omega}$ be the vector associated to the
antisymmetric operator $\Omega$, {\it i. e.} $\boldsymbol{\omega}\times
\y := \Omega(\y)$ [see  \citet{crampin}, p. 97].
Now, Equation (\ref{veloc}) can be written in the more familiar form
\begin{equation}
\v=\V+\boldsymbol{\omega}\times\left[\lambda(\rp)\right]+
\lambda(\vp)\label{veloc-p}.
\end{equation}
Hence, $\boldsymbol{\omega}$ is the usual angular velocity and reasonably
we can call $\Omega$ the angular velocity operator.

As an important application of the angular velocity operator, we are
going to derive an intrinsic expression for Coriolis theorem (see
  \citet{dede} for an unconventional derivation of that theorem
based on an interesting graphical trick, a kind of our \sello).
For that aim, we
time differentiate Equation (\ref{veloc}) and obtain
\begin{eqnarray}
\a=\dot\V+\la(\dot\vp)+\dot\Omega\left[\la(\rp)\right]+
\Omega\left[\dot\la(\rp)\right]\nn\\
+\Omega\left[\la(\vp)\right]+\dot\la(\vp)\nn\\
=\A+\la(\ap)+\dot\Omega\left[\la(\rp)\right]+
\Omega\left[\dot\la\comp\la^{-1}\comp\la(\rp)\right]\nn\\
+\Omega\left[\la(\vp)\right]+\dot\la\comp\la^{-1}\comp\la(\vp)\nn.
\end{eqnarray}
that, by using the definition of the angular velocity, can be written as
\begin{equation}
\a=\A+\la(\ap)+\dot\Omega\left[\la(\rp)\right]+
\Omega\left\{\Omega\left[\la(\rp)\right]\right\}+
2\Omega\left[\la(\vp)\right]\label{coriolis},
\end{equation}
the advertised Coriolis theorem. In this form, we see clearly
which space each quantity belongs to. Besides, the derivation
is straightforward and intrinsic. Finally, it is worth pointing
out that whenever  quantities in $K'$ are to be obtained from
those in $K$ or vice versa it suffices to use $\lambda$
and its time derivatives instead of $\Omega$. The intrinsic definition
of the angular velocity makes possible to formulate the kinematics and
dynamics of the rigid body in a completely intrinsic way.

\subsection{Geometry of the rotating disk in special relativity}
It is well known the important heuristic role played by the
rotating disk in the formulation of General Relativity
[see \citet{stachel}]. Here we approach the study
of the geometry of the rotating disk in the framework of
the Special Relativity in a
similar way as in \citet{meller}, but trying to
clarify the nature both of the problem statement and
its solution by using the intrinsic method previously
introduced.

The first problem is to give a precise and meaningful
definition of a rotating disk.
On the one hand, as in the rest of this paper, $K$
is an inertial body of reference.
As such, the corresponding space is euclidean
and provided with a set of
synchronized clocks. On the other hand, $K'$
designates the  rotating disk which is
made of a continuum of particles.

As before, we define the trace of any particle of the disk
$K'$ on $K$ at time $t$ as the point in $K$ coinciding with
it at that time. We call $\Lambda_t$ the \sello\ of $K'$ on $K$,
$$\Lambda_t : K' \rightarrow K.$$

\noindent {\bf Definition:} {\it We say that $K'$ is a rotating disk
around a point $O$ of $K$ if the trace of any particle $P'$ of $K'$
describes a uniform circular motion around $O$, with an angular velocity
$\omega$, the same for every particle.}

\vspace{0.1in}
\noindent We define a rotation operator in $K$ around the point $O$ by an angle
$\theta$ as $e^{i\theta}$,
\begin{equation}
{\r} \mapsto e^{i\theta} \r.
\end{equation}
Let  $P'$ be a fixed point in $K'$. By definition, the vector
$\overrightarrow{O\Lambda _t(P')}$ can be seen as the rotated
$\overrightarrow{O\Lambda _0(P')}$ by the angle $\omega t$, thus
\begin{equation}
\overrightarrow{O\Lambda_t(P')}=
e^{i\omega t} \overrightarrow{O\Lambda_0(P')}.
\end{equation}
The velocity of $P'$ at $t$ is
\begin{equation}
\V (P^{\prime },t)=\frac{\partial }{\partial t}\overrightarrow{O\Lambda
_{t}(P^{\prime })}=i\omega e^{i\omega t}\overrightarrow{O\Lambda
_{0}(P^{\prime })}=i\omega \overrightarrow{O\Lambda_{t}(P^{\prime })},
\end{equation}
where $i$ rotates vectors by an angle of $\pi/2$ in the positive sense.
This result can be stated {\it \`a la Euler} as the velocity of the
disk particle coinciding at $t$ with the point $P$ of $K$ given as
$\overrightarrow{OP}=\overrightarrow{O\Lambda_{t}(P^{\prime })}$, i. e.,
the trace of $P'$ at $t$. Thus
\begin{equation}
\label{velocidades}\v (P)=i\omega \overrightarrow{OP}=
\V (P^{\prime },t) \mbox{ provided } P=\Lambda_{t}(P^{\prime }).
\end{equation}

Now, we are going to construct the disk spatial metric.
For that aim we introduce a collection of inertial bodies,
comoving each with the velocity $\v (P)$.
Thus, consider the reference body $K'_P$ comoving with the disk
particle passing by $P$ at $t$.
Being at relative rest, a little patch of $K'$ around $P'$ can be
identified with a similar portion of $K'_P$, and thus the mapping
$\Lambda_t^P: K'_P\rightarrow K$ can be used for transforming the
portion of $K'$ into $K$. Therefore,
\begin{equation}
\Lambda_t^P(P'+d{\bf r'})=\Lambda_t^P(P')+\lambda^P(d{\bf r'}).\nonumber
\end{equation}
Hence, the little displacement $d {\bf r}$ at $K$ and the
corresponding $d {\bf r}'$ at the disk are related as
\begin{equation}
d{\bf r}=\Lambda_t^P(P'+d{\bf r}')-\Lambda_t^P(P')=\lambda^P(d{\bf r}')
\end{equation}
   From Theorem \ref{thlambda},
\begin{eqnarray}
d{\bf r}_{\plano}^2=d{\bf r'}_{\plano}^2 \nonumber\\
d{\bf r}_{\recta}^2=\frac{1}{\gamma^2}d{\bf r'}_{\recta}^2,
\nonumber
\end{eqnarray}
where, according to Equation (\ref{VALORGAMMA}),
\begin{equation}
\gamma=\left(1-\frac{\omega^2}{c^2}r^2\right)^{-1/2}.\label{d_gamma}
\end{equation}
Hence,
\begin{equation}
d{\bf r'}^2=d{\bf r}_{\plano}^2+\gamma^2d{\bf r}_{\recta}^2.\label{d_metric}
\end{equation}
Thus, we have obtained the local metric of the disk in terms of
the metric of its trace on $K$. This metric confers the disk a
structure of a Riemannian manifold.

We parametrize the points of the disk manifold by introducing
polar coordinates in $K$. The velocity of a point $P(r,\theta)$
of the disk, according to Equation (\ref{velocidades}), is
$\v(P)=r \omega \hat {\bf e}_\theta$. A small but
arbitrary displacement from $P$,
$d{\bf r}=dr\, {\hat {\bf e}_r}+rd\theta\, {\hat {\bf e}_\theta}$,
can be divided into parallel and perpendicular components
respect to $\v(P)$ as
\begin{equation}
d{\bf r}_{\recta}^2=r^2d\theta^2 \mbox{ and } d{\bf r}_{\plano}^2=dr^2.
\end{equation}
Then, the expression for the metric (\ref{d_metric}) in
the chosen chart is
\begin{equation}
d{\bf r'}^2=dr^2+\gamma^2 r^2 d\theta^2,
\end{equation}
where $\g$, as given by (\ref{d_gamma}), depends on the radial
coordinate $r$. Finally, the element of arc length is
\begin{equation}
dl'=\sqrt{d{\bf r'}^2}=\sqrt{dr^2+\gamma^2 r^2 d\theta^2}.
\end{equation}
Now, it is immediate to obtain the disk radius length from
the element of arc by taking $d\theta=0$:
$$
R'=\int_0^R \sqrt{dr^2+\gamma^2 r^2 0^2}=R
$$
The disk rim length is obtained by integration of the arc
element along $r=R$:
$$
l'=\int_0^{2\pi}\sqrt{0^2+\gamma^2 r^2 d\theta^2}=\frac{2\pi R}
{\sqrt{1-\frac{\omega^2}{c^2}r^2}}.
$$

\section{Conclusions}
We have developed an intrinsic method for deriving space and time
transformations. For that purpose, following a suggestion by Einstein, we
associated a different space of reference to every reference body. In
order to connect the space and time measurements in the spaces we introduced
the  \sello, a simple concept fitted with a direct physical
meaning, which makes possible to express with a greater clarity
several ideas of the special relativity.

The intrinsic method shows some advantages we are going to summarize:

\begin{enumerate}
\item The space and time transformations were derived using no coordinate
systems. Therefore, a clear distinction is now possible between a coordinate
system and a body of reference.
\item The \sello\ allows a sharp definition for the geometry of a moving
body:
It is nothing but the geometry of its trace. A reference space not
equipped with clocks is an incomplete system to
register events. Nevertheless, as seen in the discussion of the
rotating disk, we were able to obtain its geometry by using
the \sello\ as it does nor require any time measurement in that space.
\item The intrinsic method does not depend in any way on a notion
of parallelism between both spaces of reference.
Rather, the trace allows us to define that
parallelism in an operational way.  Note that, in the non--intrinsic
derivations of the spacetime transformations,
the coordinate axes of both reference frames are assumed
parallel to each other. But, here we  have not argued,
for example, that $\u$ and $\up$ are parallel because
they belong to different spaces. We affirm, instead, that
$\u$ and $\lambda(\up)$ are parallel.
\item Even though in the Galilean space and time one might identify the
different spaces of reference, the lesson our derivation of Coriolis Theorem
teaches us is that the \sello\ between separated spaces is useful even in 
these
cases, in which one could identify them.
\end{enumerate}

\section*{Acknowledgements}
We are indebted to Professors J. Casahorr\'an, A. Vi\~na and J.
Fern\'andez-N\'u\~nez
for helpful discussions and kind advice.
We are particularly grateful to Professor T. Matolcsi for sending us
a copy of his
most valuable book referenced as \citet{matolcsi}.

\vfill
\end{document}